\documentstyle[12pt,epsf]{article}
\def\la{\mathrel{\mathpalette\fun <}}

\def\fun#1#2{\lower3.6pt\vbox{\baselineskip0pt\lineskip.9pt
\ialign{$\mathsurround=0pt#1\hfil##\hfil$\crcr#2\crcr\sim\crcr}}}
\title{BOUNDS ON THE NEW GENERATIONS}
\author{V.Novikov\\
ITEP, Moscow 117259, Russia}
\date{}
\begin{document}
\maketitle

\vspace{6cm}

\begin{abstract}

We consider the bounds for the values of higgs mass $m_H$ and the
mass of the extra quarks and leptons $m_{extra}$ derived from the
stability of electroweak vacuum and from the absence of Landau pole
in Higgs potential. We find that in the case of the absence of new
physics up to the GUT scale the bounds on masses of 4th generation
are so strong that one can hope to discover it at LEP2.

\end{abstract}

\newpage

Last year CDF and D0 collaborations announced about observation of
$t$-quark -- the last missing member of the third generation. The
first one, $\tau$-lepton, was discovered 20 years ago, i.e. at the
time when practically nobody begged for a new quark-lepton generations.
From LEP1 experimental results we have learnt already that the
sequence of the generations with light neutrino is completed with
$N_f  \mbox{\rm(light neutrino)}=3$. As for the theory, nobody
suggested yet a good explanation of the fact that $N_f  = 3$, or of
the fact that neutral leptons are massless. So it is natural to look
for  new sequential generations $\left( \begin{array}{c}T
\\ B \end{array}\right)_i$ and $\left(\begin{array}{c}N \\ E
\end{array} \right)_i$ ($i=4,5,...$) with heavy neutral lepton $N$:
$m_N > \frac{1}{2} m_Z$.

There exist a few direct and indirect bounds on the masses of these
new sequential generations.

\underline{Experimental bounds.}

From LEP1.5 searches it follows that
\cite{1}:
\begin{equation}
m_E > 62 GeV
\label{1}
\end{equation}
$$
m_N > 58.9 GeV
$$

Bounds on the masses of new quarks depend crucially on the
assumptions about their decay. For example, for quarks, stable enough
to leave the detector, the limit is \cite{2}:
\begin{equation}
m_{T,B} > 139 GeV
\label{2}
\end{equation}

For unstable quarks this number is lower. It is rather safe to say
that at the moment absolute bound for quarks is somewhere near 100
GeV. These are, so to say, direct bounds. They demonstrate that new
generations should be rather heavy.

\underline{Theoretical bounds.}

The study of radiative corrections for LEP1 data produces additional
indirect bounds \cite{3}. The main facts are that the radiative
corrections for LEP1 observables are unexpectedly small (of the order
of $10^{-3}$) and that the contributions of heavy quarks and leptons
into low-energy observables (e.g. into LEP1 observables) are finite
even for heavy mass goes to infinity (the absence of decoupling of
chiral matter). So to protect the successful description of the
precision data in the framework of the Standard Model from
undesirable new contributions we have to impose some bounds on new
physics.

First of all following to the Veltman's arguments we
conclude that the masses of isopartner ($m_T$ and $m_B$,
$m_N$ and $m_L$) should be almost degenerate. Indeed for the case of
very different masses $m_T$ and $m_B$ we
effectively  get a violation of SU(2)$_L$
symmetry at low energy and, as a result, a large low-energy
loop corrections of the order of
\begin{equation}
\left(\frac{\alpha_{W,Z}}{\pi}\right)
\frac{m_T^2 -m_B^2}{m_Z^2} \sim 10^{-2} \frac{m_T^2 -m_B^2}{m_Z^2}
\label{3}
\end{equation}
where $\alpha_{W,Z}$ are the weak coupling constants. Since there is
no room for large corrections to the Standard Model values for
electroweak observables, the masses of the new quarks and leptons in
SU(2) doublets should be approximately degenerate, i.e.
$$
\frac{|m_T^2 -m_B^2|}{m_Z^2} \la 1 \;\; , \;\; \frac{|m_E^2
-m_N^2|}{m_Z^2} \la 1 \;\; .
$$

So hereafter we assume, that
\begin{eqnarray}
m_T\simeq m_B \simeq m_Q \nonumber \\
 \\
m_E \simeq m_N \simeq m_L \nonumber
\label{4}
\end{eqnarray}
(To reduce the number of parameters we assume also that $m_Q \simeq
m_L =m_{extra}$).

The study of the radiative corrections with degenerate and heavy
masses is a little bit more subtle matter than the previous case. The
result is that starting from $m_{extra}\simeq 60$ GeV (LEP 1.5
bounds) the additional contribution to LEP observables due to extra
generation rather weakly depend on $m_{extra}$ and the effect of one
new generation can be compensated by increasing
of the fitted value of
$m_{top}$ by 10 GeV \cite{3}. The experimental accuracy for $m_{top}$
is of the order of 20 GeV, the accuracy of the fitted value of
$m_{top}$  from LEP1 data is of the order of 10 GeV.
So direct measurements and precision LEP data allow to
have one or two new sequential heavy generations.

But there is another way to attack the problem and to get the
\underline{upper}
bounds both on the masses and on the number of new generations
\cite{4}.  The point is that heavy fermions change the vacuum energy
(i.e. the higgs potential) through radiative correction \cite{5,6}.
Roughly speaking the heavy fermions  give  a negative contribution
into higgs potential due to Fermi statistic. In one loop
approximation this contribution looks like
$$
\Delta V(\phi) =  -\frac{1}{16\pi^2} \sum_{f=Q,L}(\frac{N_c^f
m_f^4}{\eta^4}) \phi^4 \ln \frac{\phi^2}{\eta^2}\;\;,
{\rm where}\;\; N_c^Q = 3\;\;, \;\; N_c^L = 1\;\;.
$$

So if heavy fermions  are not accompanied by scalars (or by vectors)
to compensate this negative contribution, the electroweak vacuum
$\phi = \eta$ becomes unstable for large value of $\phi$.
There are rather subtle bounds on the values of the parameters for
selfconsistent theory. For the Standard Model the experimental
value of  top mass $m_t = 180$ GeV is very close to the edge of the
region for allowed values of heavy mass. So for SM the vacuum is
already
oversaturated by heavy top alone and there is practically no room for
new heavy fermions (without new scalars).

Technically this problem reduces to the solution of renormgroup
equations for running coupling constants.
With the account of radiative corrections the renormalization group
improved higgs potential can be presented in the following form \cite
{6}:
\begin{equation}
V[\phi] = -\frac{1}{2}\mu^2(t) [G(t)\phi]^2 +\frac{1}{4}\lambda(t)
[G(t)\phi]^4
\label{5}
\end{equation}
where $\phi$ is the higgs field, $t=\ln(\phi/\eta)$, $\eta = 246$ GeV
is v.e.v. of $\phi$ in electroweak vacuum, $\mu(t)$ and $\lambda(t)$
are running constants and $G(t)$ is determined by anomalous dimension
of the field $\phi$.

For $\phi\sim\eta$ the initial values of $\mu$ and $\lambda$ govern
$V(\phi)$ behaviour, while for $\phi \gg \eta$ the radiative
corrections to $\lambda(t)$ becomes essential. For $\phi \gg \eta$
the term $\sim \phi^2$ is negligible and only the behaviour of
$\lambda(t)$ is crucial for the selfconsistency of the theory.

If $\lambda(t)$ becomes negative at some value $\phi_0$, then the
$V(\phi)$ minimum $\phi =\eta$, that corresponds to our electroweak
vacuum, becomes unstable. So if we wish to live in the stable 
Universe, we have to avoid such values of the
parameters that lead to vacuum instability.

On the other hand for large higgs mass $m_H$
the coupling constant $\lambda(t)$ becomes infinite
at some value  $t_0 =\ln \phi_0/\eta$ .
This is so called Landau pole in $\lambda(t)$. (Such behaviour for
selfinteraction of scalar and for Yukawa coupling was first
discovered in perturbation theory in 50th. Later it was confirmed by
computer calculations and by more rigorous arguments). So to avoid
strong interaction in Higgs sector we demand the absence of Landau
poles in $\lambda(t)$.

So our restriction is:  $0 < \lambda{(t)} < \infty$
up to the scale $\Lambda$ at which new physics begins. In order to get
$\lambda(t)$ behavior one needs the renormalization group equations
which determine the behavior of $\lambda$, Yukawa coupling constants
of $t$-quark and new quarks and leptons  with Higgs doublet and gauge
coupling constants.  These equations can be found in literature
\cite{7}.  Let us present here the renormalization group
equation for the running value $\lambda(t)$ in a theory with $N$ generations
of heavy fermions with the degenerate masses:
\begin{eqnarray}
\frac{d\lambda}{dt} &=& \frac{3}{2\pi^2}
\lambda^2 + \frac{\lambda}{4\pi^2} [3g^2_t +
3N(g^2_T + g^2_B) + N(g^2_E +
g^2_N)] - \nonumber \\
&-& \frac{1}{8\pi^2}[3g^4_t + 3N(g^4_T + g^4_B) +
N(g^4_E + g^4_N)] -\nonumber \\
&-& \frac{3}{16\pi^2} \lambda(3g^2 +
g'^2) + \frac{9}{384\pi^2} g'^4 + \frac{9}{192\pi^2} g^2g'^2 +
\frac{27}{384\pi^2} g^4\;\;,
\label{6}
\end{eqnarray}
where $g$ and $g'$ are SU(2) and U(1) coupling constants and
constants $g_i$ determine the masses of the corresponding fermions
by  the formula:  $m_i = \frac{g_i(0)\eta}{\sqrt{2}}$,
$g_t(0) = 1.035$ corresponds to $m_t = 180$ GeV.
We should add to equation (\ref{6}) the renormgroup equations
for $g, g'$, SU(3) coupling  constant $g_3$ and $g_i$ and
numerically integrate the system of the coupled differential
equations.
Results of the integration for the case of one heavy generation are
presented in Fig. 1.
\begin{figure}
\epsfbox{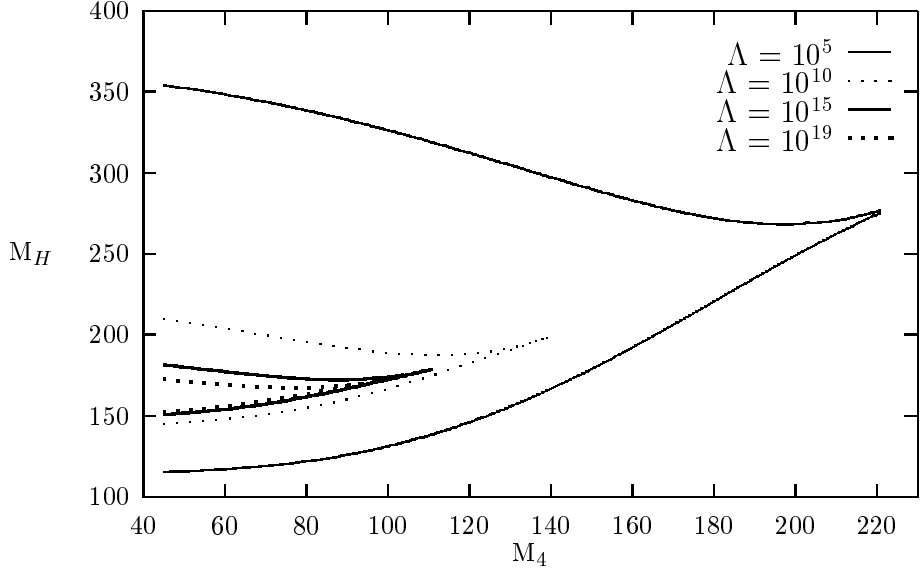}
\caption{ Allowed values of $M_H$ and $M_4$ lie between two curves:
a. solid for $\Lambda = 10^5$ GeV;
b. thin dotted for $\Lambda = 10^{10}$ GeV;
c. thick solid for $\Lambda = 10^{15}$ GeV;
d. thick dotted for $\Lambda = 10^{19}$ GeV. }
\end{figure}
If the value of $m_4$ is small, then we approximately get the
allowed interval
of the values of $m_H$ for $m_t = 180$ GeV
in the Standard  Model, well-known from literature
\cite{6}. For all values of
ultraviolet cutoff
$\Lambda$ the low lines which represent $\lambda = 0$ stability
bound go up with growing of $m_4$.  The physical reason for such
behavior is clear -- the third term in equation (\ref{6}) becomes
larger and a heavier Higgs is required to get a positive potential
for heavier fermions. The upper lines which represent Landau pole
bound are governed by the first term in (\ref{6}) and are almost
constant for $\Lambda > 10^{10}$ GeV.  However, for $\Lambda = 10^5$
GeV new phenomena occur -- the second term in (\ref{6}) becomes
essential and an upper curve goes down for increasing $m_4$. So
the allowed interval of $m_4$ values shrinks.

From Fig. 1 we see that for traditional Grand Unified Theories,
for which new physics does appear only at $\Lambda \sim 10^{15}$ GeV,
the bounds for the mass of 4th generation is very low, $m_4 \leq 100$
GeV. This is  a region available for LEP2 research. As for higgs
mass, in this case it is fixed between 160 and 180 GeV.

If we decrease the bounds for new physics down to $\Lambda = 10^{10}$
GeV, the bounds for $m_4$ increases up to $m_4 < 140$ GeV, i.e. that
is exactly CDF bound for stable quarks (see eq. (\ref{2}).

The introduction of the additional heavy generations will
restrict the allowed values of $m_{extra}$ and $m_{Higgs}$. For
example, for $N=3$ and $\Lambda = 10^{10}$ GeV only $m_{extra} <
90$ GeV is allowed (see Fig. 2).
\begin{figure}
\epsfbox{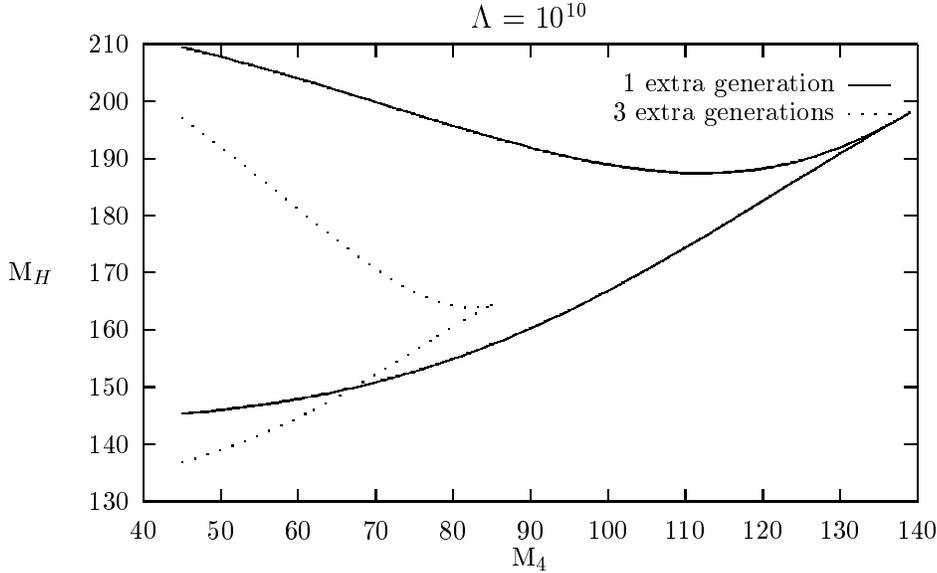}
\caption{ Allowed values of $M_H$ and $M_4$ for 1 and 3 extra
generations for $\Lambda = 10^{10}$ GeV. }
\end{figure}
In our analysis we use the one loop renormalization group potential.
It is not difficult to take into account the second loop as well. In
this case we expect the change of the bounds by approximately 10
GeV, i.e. not very drastically.

\underline{Conclusions.}

We see that in the SM we are allowed to add only one (or two) new
sequential generation with rather low masses. They could be
discovered at LEP2.

I am grateful to H.Nielsen, A.Novikov, and M.Vysotsky for a pleasure
to discuss the subject and to be coauthor of the paper \cite{4}.

\end{document}